\documentclass{elsart}
\usepackage{graphicx}

\usepackage{amssymb}
\usepackage{amsmath}

\begin{document}

\begin{frontmatter}



\title{Role of transcriptional bursts in cellular oscillations}


\author{N. Almeira}
\address{Facultad de Matem{\'a}tica, Astronom{\'\i}a, F{\'\i}sica y Computaci{\'o}n, Universidad Nacional de
C{\'o}rdoba, Ciudad Universitaria, C{\'o}rdoba, Argentina.\\
Instituto de F{\'\i}sica Enrique Gaviola (IFEG-ConICET), Ciudad Universitaria, 5000 C{\'o}rdoba,
Argentina.}

\author{S. Risau-Gusman}
\address{Centro At{\'o}mico Bariloche,
8400 S. C. de Bariloche, Argentina \\
Consejo Nacional de Investigaciones Cient{\'\i}ficas y
T{\'e}cnicas, Argentina}
\begin{abstract}
Genetic oscillators are present in the cells of many organisms and control several biological processes. The common feature of such oscillators is the presence of a protein which represses the transcription of its own gene. Recently, it has been shown that for many genes transcription is not a continuous process, but that it proceeds in bursts. We study here the  relationship between bursty transcription and the robustness of protein oscillations. We concentrate on the temporal profile of mRNA production by studying regimes where this profile changes but the amount of mRNA produced is kept fixed. For systems with different degrees of cooperativity we show that in general bursts are associated with more robust oscillations, but when they are too short and intense they can have the opposite effect. In other words, we show that, in terms of the regularity of the oscillations generated, there is an optimal value for the intensity of the bursts.
\end{abstract}

\end{frontmatter}

\section{Introduction}
\label{Intro}

Given the relatively low copy number of molecules involved, stochasticity usually plays an important role in the cell cycle. Because only one or two copies of each gene are present in a cell, the production of messenger mRNA (also called gene expression) is a process where stochastic effects are particularly relevant~\cite{Raj}. Experimental studies carried out on several organisms, ranging from yeast~\cite{Becskei} and {\it E. coli}~\cite{Elowitz} to mammals~\cite{Raj2,Suter}, have found that the number of mRNA molecules presents large variations from cell to cell. Moreover, it has been shown that in general gene expression proceeds in short but intense bursts followed by relatively long periods during which the gene is `silent'. However, it is not yet clear whether this bursty transcriptional dynamics is governed by processes acting on the whole genome or whether these processes are gene-specific~\cite{Sanchez}.

The first question that arises is what can be the benefits (if any) of having large levels of noise in gene expression~\cite{Raj}. It has been shown that stochastic expression of a very specific gene is necessary for the development of the retinal mosaic that is so characteristic of the fruit fly (Drosophila)~\cite {Wernet}. One early example of mathematical modelling has suggested that stochastic gene expression can also underlie the phenotypic variations that are observed in some colonies of both eukaryotic and prokaryotic cells~\cite{McAdams}. In turn, other mathematical models~\cite{Kussel} have shown that such phenotypic variability could confer an adaptive advantage in fluctuating environments. This was later confirmed by experiments with yeast strains~\cite{Acar}.

Noise in gene expression induces large fluctuations in the abundances of the proteins encoded. For most proteins, however, there is a well defined steady state about which this fluctuation occurs. But there are some proteins whose abundance is known to have a cyclical variation throughout the day. The best example are the proteins involved in the circadian clock~\cite{Panda}, but there are other proteins whose abundance oscillates with shorter, ultradian, periods (see e.g.~\cite{BarOr}). The basic mechanism of these oscillators is a feedback loop involving one or more proteins that repress the transcription of their own genes. Bursty transcription seems to be the dominant form of gene expression (at least for humans~\cite{Dar}) and it has recently been shown that this may also be the case for circadian genes~\cite{Suter,Ono}.

One important difference between circadian and non circadian genes is that in the former the bursts in transcription can be caused by the very protein that the gene encodes. The relationship between protein abundance and transcriptional bursting is thus much less straightforward. The circadian clock is composed of many cellular oscillators, and it controls many behaviours. As a consequence, the cellular clocks should be as accurate as possible. It is then natural to ask what is the relationship between the fundamental stochasticity of transcriptional bursting and the regularity of protein oscillations and, moreover, whether it imposes any fundamental limit on these oscillations. These are the questions that we address in this paper. 

We study the stochastic version of a simple genetic oscillator with one feedback loop for a protein that can pass through two different states. In order to study the effect of cooperativity in the repression of the gene, we consider systems with three different degrees of cooperativity. In section~\ref{basic} we present the stochastic model and the deterministic equations associated with it. In section~\ref{characterization} we give a quantification of the quality of oscillations and relate it to the amount of bursting. Section \ref{shorts} provides a simplified theoretical treatment, for a better understanding of the results given in the previous section. In the last section we summarize and discuss our main results.

\section{Model of a genetic oscillator}
\label{basic}

We consider a genetic oscillator composed by a protein, its messenger RNA and the gene that expresses it. We assume that, when it is not being repressed, the gene is in the active state (noted as $D_0$). In other words, we assume that the gene is always associated with its activator. This models the fact that, in some circadian oscillators, the activator of the gene is constitutively expressed in the cell~\cite{Houl}. When active, the gene `produces' mRNA ($M$) at a rate $k_1$. We assume that the protein passes through two states before being degraded. Translation takes place at a rate $k_3$, generating the first state of the protein ($P_1$). This is then converted into the second state of the protein ($P_2$) at a rate $k_4$. This models the phosphorylations that circadian proteins are known to undergo~\cite{Panda}, or its entrance to the nucleus. When in the second state, the protein closes the feedback loop by repressing the activator, thus turning off the gene ($R$).

In most deterministic circadian models it is assumed that there is some degree of cooperativity in the repression of the activator by $P_2$~\cite{Goldbeter}. This is usually modelled by introducing a Hill term in the differential equation for $M$. In our stochastic model cooperativity is enforced by assuming that $n$ copy molecules of $P_2$ are needed to repress the activator. Thus, the gene passes through $n$ different states ($D_i$, $i=0, \dots, n-1$) before being completely repressed. For simplicity we assume that the rate of production of mRNA is the same in all active states. In this paper we have studied the cases of $n=1$ (no cooperativity), $n=2$, and $n=3$. In the following all the equations will be written for the case $n=3$ but the modifications necessary for the other cases are straightforward.

We have also assumed that the degradation of the protein is mediated by an enzyme ($E$), whose abundance is assumed to be constitutive and given by $E_0$. The protein and the enzyme form a complex $C$ which degrades the protein at rate $k_6$, thus freeing a copy of the enzyme. This is a gross simplification of the complex degradation paths of a protein, but at least it allows us to model the saturation of those paths. Furthermore, it has been shown that degradation terms of this form are in many cases necessary to have oscillations in a dynamical system~\cite{Kurosawa}. For simplicity, and also because the small copy number of mRNA is unlikely to saturate its degradation paths, we have assumed that mRNA is degraded at a fixed rate $k_2$. 

For $n=3$, the reactions that take place in our stochastic model are:

\begin{minipage}[t]{0.5\textwidth}
\begin{align} 
D_0 & \overset{k_1}{\rightarrow} D_0 + M, \nonumber \\
D_1 & \overset{k_1}{\rightarrow} D_1 + M, \nonumber \\
D_2 & \overset{k_1}{\rightarrow} D_2 + M, \nonumber \\
M & \overset{k_2}{\rightarrow} \emptyset, \nonumber \\
M & \overset{k_3}{\rightarrow} M + P_1, \nonumber \\
P_1 & \overset{k_4}{\rightarrow} P_2, \nonumber 
\end{align}
\end{minipage}
\begin{minipage}[t]{0.46\textwidth}
\begin{align}
P_2 + E & \overset{k_5}{\underset{k_{-5}}{\rightleftharpoons}} C \overset{k_6}{\rightarrow} E,\nonumber \\
D_0 + P_2 & \overset{k_7}{\underset{k_{-7}}{\rightleftharpoons}}  D_1,\nonumber \\
D_1 + P_2 & \overset{k_8}{\underset{k_{-8}}{\rightleftharpoons}}  D_2,\nonumber \\
D_2 + P_2 & \overset{k_{off}}{\underset{k_{on}}{\rightleftharpoons}}  R. \nonumber
\end{align}
\end{minipage}

To obtain the system for $n=2$ the reactions producing $D_2$ are eliminated and $D_2$ is replaced by $D_1$. A similar modification regarding $D_1$ must be performed to obtain the system for $n=1$. In the following the expressions will be given for the case $n=3$, unless otherwise noted. The generalizations for lower values of cooperativity are straightforward. The simulations for this system were performed using the Gillespie algorithm~\cite{Gillespie}. We tested four different parameter sets for each value of $n$.

The evolution of the averages over the stochasticity is given by the following set of equations:
\begin{eqnarray} 
\dot{D_0} &=& -k_{7} D_0 P_2 + k_{-7} D_1,  \nonumber \\
\dot{D_1} &=& k_{7} D_0 P_2 - k_{-7} D_1- k_8 D_1 P_2 + k_{-8} D_2, \nonumber  \\
\dot{D_2} &=& k_8 D_1 P_2 - k_{-8} D_2 - k_{off} D_2 P_2 + k_{on} R, \nonumber \\
\dot{M} &=& k_1 (D_0+D_1+D_2) - k_2 M, \nonumber \\
\dot{P_1} &=& k_3 M - k_4 P_1, \nonumber \\
\dot{P_2} &=& k_4 P_1 - k_5 P_2 E + k_{-5} C - k_{7} D_0 P_2 + k_{-7} D_1 + \nonumber \\ &&- k_8 D_1 P_2 + k_{-8} D_2 - k_{off} D_2 P_2 + k_{on} R,  \nonumber \\
\dot{E} &=& -k_5 P_2 E + k_{-5} C + k_6 C, \nonumber \\
\dot{C} &=& k_5 P_2 E - k_{-5} C - k_6 C,  \nonumber \\
\dot{R} &=& k_{off} D_2 P_2 - k_{on} R,
\label{eq.detailed}
\end{eqnarray}
with the initial condition at $t = 0$,
\begin{equation}
(D_0, D_1, D_2, M, P_1, P_2, E, C, R) = (D,0,0, 0, 0, 0, E_0, 0, 0).
\end{equation}
All the variables in these equations represent volumetric concentrations. The fact that the amount of enzyme and DNA remain constant induces the constraints $E_0 = E(t) + C(t)$ and $D = D_0(t) + D_1(t) + D_2(t) + R(t)$. In each cell there are only one or two copies of each gene but, for the sake of simplicity, we will assume in the following that $D=1$. Notice that this forces us to assume that the unit volume used in the volumetric concentrations is the volume of the whole cell.

It can be shown that the system given by Eqs.~(\ref{eq.detailed}) has always a single fixed point (and the same happens when the degree of cooperativity is lower). At the fixed point, the variables satisfy the equations:
\begin{eqnarray}
D_0^*&=&\frac{1}{1+\alpha_7 P_2^*+\alpha_7 \alpha_8 (P_2^*)^2+\alpha_7 \alpha_8 \alpha_o (P_2^*)^3}  \nonumber \\
D_1^*&=&\alpha_7 D_0^* P_2^*  \nonumber \\
D_2^*&=&\alpha_7 \alpha_8 D_0^* (P_2^*)^2  \nonumber \\
M^*&=&\frac{k_1(D_0^*+D_1^*+D_2^*)}{k_2}  \nonumber \\
P_1^*&=&\frac{k_1 k_3 (D_0^*+D_1^*+D_2^*)}{k_2 k_4}  \nonumber \\
E^*&=&E_0-\frac{k_4 P_1^*}{k_6}  \nonumber \\
E_0 &=& \frac{E_0}{P_2^*+x} +\frac{k_1 k_3(1+\alpha_7 P_2^*+\alpha_7 \alpha_8 (P_2^*)^2)}{k_6 k_2(1+\alpha_7 P_2^*+\alpha_7 \alpha_8 (P_2^*)^2+\alpha_7 \alpha_8 \alpha_o (P_2^*)^3)}
\label{eq.determ}
\end{eqnarray}

\noindent where $\alpha_i=k_i/k_{-i}$, $\alpha_o=k_{off}/k_{on}$ and $x=(k_6+k_{-5})/k_5$.

Oscillatory solutions appear when the fixed point becomes unstable, usually at a supercritical Hopf bifurcation. When the system parameters lie in the region where the system displays oscillations, and they are close to the bifurcation, the time average of each system variable is very close to the value of that variable at the fixed point (because the limit cycle is very close to the unstable fixed point). However, this is not necessarily the case when the system is in the oscillatory regime but far from the Hopf bifurcation.

If we take the time average of Eqs.~(\ref{eq.detailed}), when times are long enough the average of all left sides will vanish (because the system can be assumed to be in the limit cycle). In principle, the average value of each variable cannot be obtained from this equation system because of the presence of nonlinear terms. However, we can calculate some average values if we assume that during most of the oscillation the degradation path of the protein is saturated. In this regime we have $\langle E \rangle \ll E_0$, and averaging Eqs.~(\ref{eq.detailed}) over time, we obtain that the time average of $M$ can be well approximated by:
\begin{equation}
\langle M \rangle \approx \frac{k_6 E_0}{k_3}
\label{eq.M}
\end{equation}
\noindent whereas for the active state of the gene we have 
\begin{equation}
\langle D_0+D_1+D_2 \rangle \approx \frac{k_2 k_6 E_0}{k_3 k_1} 
\label{eq.D}
\end{equation}

Note that in the stochastic system $D_0+D_1+D_2$ can only switch between $0$ and $1$ (assuming there is only one copy of the gene), and thus $p_{on} \equiv \langle D_0+D_1+D_2 \rangle$ represents the fraction of time that the gene spends in the active state. One possible caveat is that the relationship given in Eq.5 applies strictly only to the time average of the ensemble average (i.e. the average over simulations) of the fraction of time that the gene is active. But we have found that the difference between $p_{on}$ and the deterministic value is very small in every single simulation.

\section{Characterization of the oscillations}
\label{characterization}

We are interested here in studying the effects of bursty transcription on the dynamics of the genetic oscillator. In other words, we want to compare systems where the time average of the mRNA expressed by the gene is the {\it same}, but with different temporal profiles. To avoid a systematic search in the parameter space of the stochastic system, we turn to the deterministic equations that govern the average (over the stochasticity). We look first for sets of parameters that are within a biologically realistic range and such that the time average of the number of copy molecules of mRNA is relatively low. To modulate the time profile of mRNA expression we then choose $k_{off}$ and $k_1$ as control parameters. Furthermore, given $k_{off}$, $k_1$ is given the value

\begin{equation}
k_1=f(k_{off})= \frac{k_2 M^*}{D} \left(1+ \frac{\alpha_7 \alpha_8 (k_{on}/k_{off}) (P_2^*)^3}{1+\alpha_7 P_2^*+\alpha_7 \alpha_8 (P_2^*)^2} \right)
\label{eq.koff}
\end{equation}

These parameters control the length and the intensity of the transcriptional bursts. In particular, Eq. (\ref{eq.D}) shows that they can be directly related to the fraction of time that the gene spends in the bursting state. 

\begin{figure} 
\centerline{\includegraphics[width=\columnwidth,clip=true]{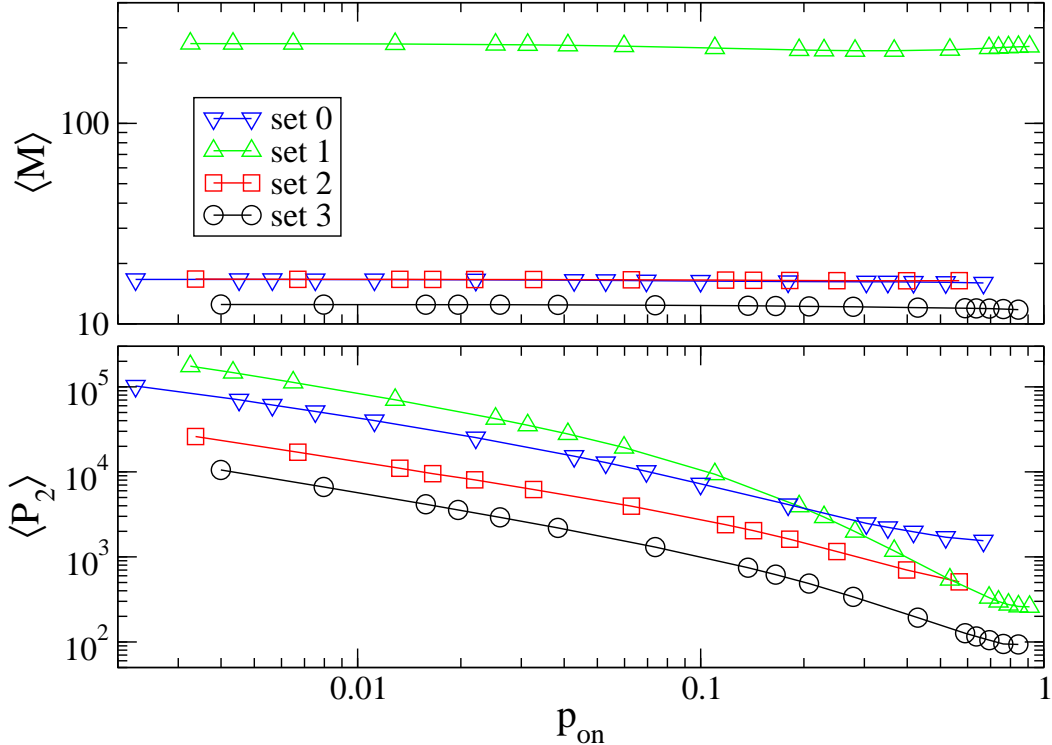}}
\caption{Time averages of the amount of mRNA (upper panel) and the second state of the protein, $P_2$, (lower panel) in the stochastic system with two levels of cooperativity, as a function of $p_{on}$, the fraction of time that the gene spends in the active state. $k_1$ satisfies $k_1=\frac{k_2 k_6 E_0}{k_3 p_{on}}$, and $k_{off}=f(k_1)$ (Eq. (\ref{eq.koff})). The rest of the parameters belong to 4 different parameter sets, which are fully specified in the Supplementary Material.
} \label{figure1}
\end{figure}

Using Eqs. (\ref{eq.determ}) it is straightforward to show that if $k_1$ and $k_{off}$ are modified using Eq. (\ref{eq.koff}) the values at the fixed point of $M^*$, $P_1^*$, $P_2^*$ and $E^*$ do not change. Thus, if we choose a parameter set such that $\langle E \rangle \approx E^* \ll E_0$ (i.e. the degradation path of the protein is saturated during most of the oscillation), and then $k_1$ and $k_{off}$ are modified using Eqs. (\ref{eq.determ}), it can be expected that this approximation will continue to hold. This, in turn implies that the time average of $M$ remains almost constant. Fig. \ref{figure1} shows that for the parameter sets analyzed this expectation is justified.

\begin{figure} 
\centerline{\includegraphics[width=\columnwidth,clip=true]{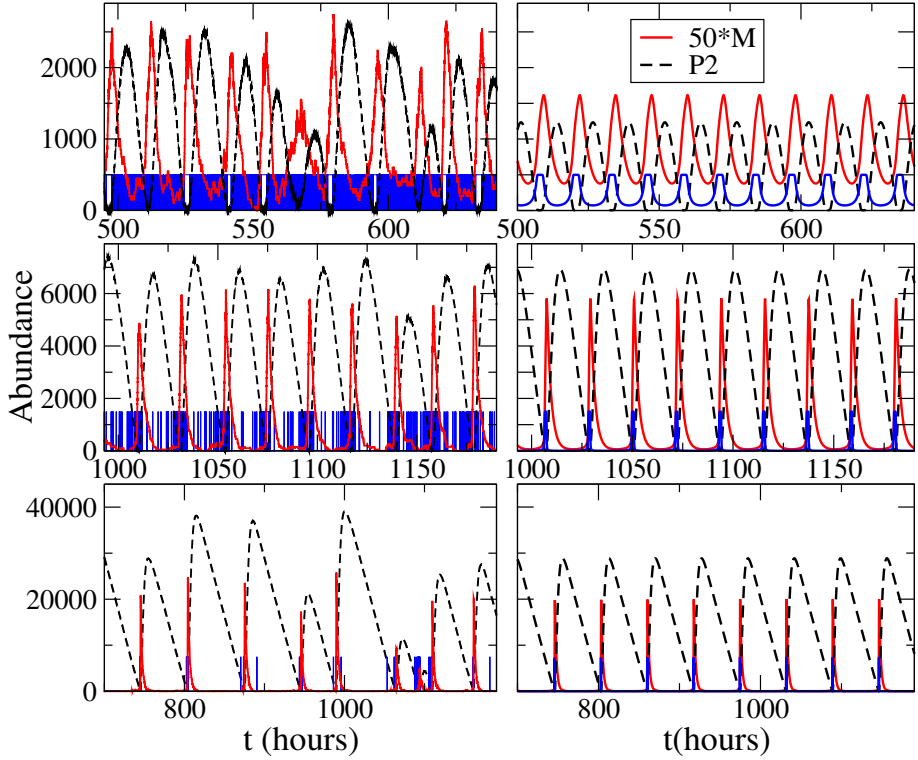}}
\caption{Abundance of protein ($P_2$) and mRNA ($M$) as a function of time, for one simulation of the stochastic genetic oscillator (left column) and for its determnistic counterparts (right column), for three different values of $p_{on}$ and $n=3$. The parameters used correspond to set 2. The blue spikes in the left panels represent the moments when the gene is in the active state. The blue curves in the right panels give the abundance of $D_0+D_1+D_2$. These abundances have been rescaled in order to make them visible (the maximum of the abundance is always $1$). Upper panels: $p_{on}=0.398$, middle panels: $p_{on}=0.063$, lower panels: $p_{on}=0.007$.} \label{figure2}
\end{figure}

On the other hand, when the gene spends less time in the active state (what leads to more intense bursts) the amount of protein produced is increased (see Fig. \ref{figure1}). This happens because the temporal profile of mRNA production becomes very sharply peaked as $p_{on}$ is decreased (see Fig. \ref{figure2}). Note that this does not change the {\it average} amount of mRNA that is produced. This, in turn, implies that a large amount of protein is produced within a short time, and these proteins last much longer because the degradation path is strongly saturated.

\begin{figure} 
\centerline{\includegraphics[width=\columnwidth,clip=true]{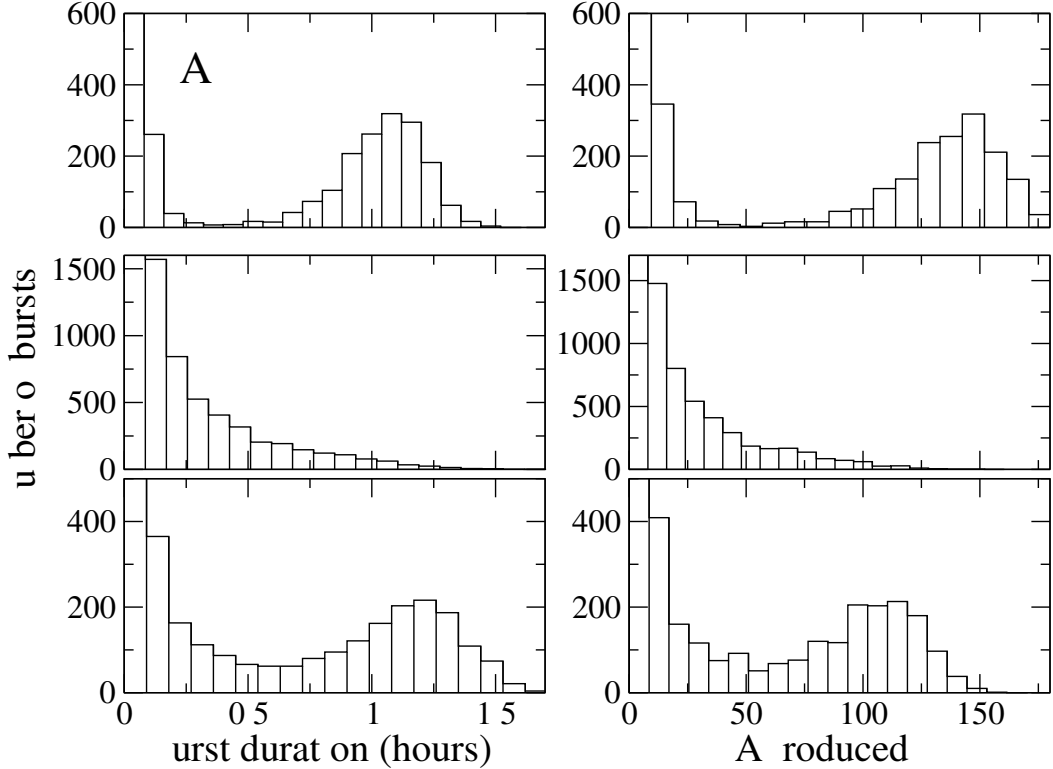}}
\caption{Histograms of the bursts generated in simulations of 35000 system hours, for systems with $n=3$ (panels A and B), and $n=1$ (panels C and D), as a function of the duration of the bursts (left column) or of the number of mRNA molecules produced in the burst (right column). The parameters used correspond to set 2 (see Supplementary Material). Panels E and F show the histograms for bursts obtained by joining bursts separated by less than $20 \%$ of the duration of the preceding burst, for the system with $n=1$ .
} \label{figure3}
\end{figure}

As Fig. \ref{figure2} illustrates, both the time of occurrence of bursts and their duration are stochastic processes. It is natural to associate long bursts with large excursions of protein levels. On the one hand because such bursts tend to produce the relatively large amounts of mRNA necessary to make protein levels grow, and on the other hand because long bursts are much more likely to be produced when protein levels are very low. Furthermore, if the oscillations are more or less regular, the distribution of burst durations displays a mode at relatively large time values that are related to the oscillations of the protein. Let us consider, as an example, the system with $n=3$, and the parameters in set 2 and $k_{off}=2$. The average distance between minima of $P_2$ is $21.6$ hours, and the fraction of time that the gene spends in the active state is $p_{on}=0.063$, that is, approximately $1.36$ hours per cycle. Fig. \ref{figure3} shows that the mode of the distribution is at $1.1$ hours, what implies that in most cases a single long burst is responsible for most of the activity of the gene between minima of abundance of the protein.

Interestingly, the presence of such a mode is highly dependent on the cooperativity of the system. Panels C and D of Fig. \ref{figure3} show that even when oscillations are very regular, the distribution of bursts duration does not have a mode and is in fact very similar to an exponential distribution (which is one of the most frequent distributions for transcriptional bursts associated to proteins that are constitutively expressed~\cite{Raj2,Chubb}). What happens in this case is that the binding or unbinding of only one molecule of the protein is enough to repress or activate, respectively, the gene, what makes long active periods much less frequent. However, when many molecules of protein are present, these processes are rather fast. As a consequence, the gene is only inactive for a very brief time. In some sense, we could say that in this case the long bursts of the cooperative case are broken down into a series of shorter bursts. This picture is confirmed by panels E and F of Fig. \ref{figure3}. There we show the distributions that are obtained when all the bursts that are separated by short times are combined in a single burst. In the case displayed in the figure times are defined as short when the are less than $20 \%$ of the duration of the preceding burst.

\begin{figure} 
\centerline{\includegraphics[width=\columnwidth,clip=true]{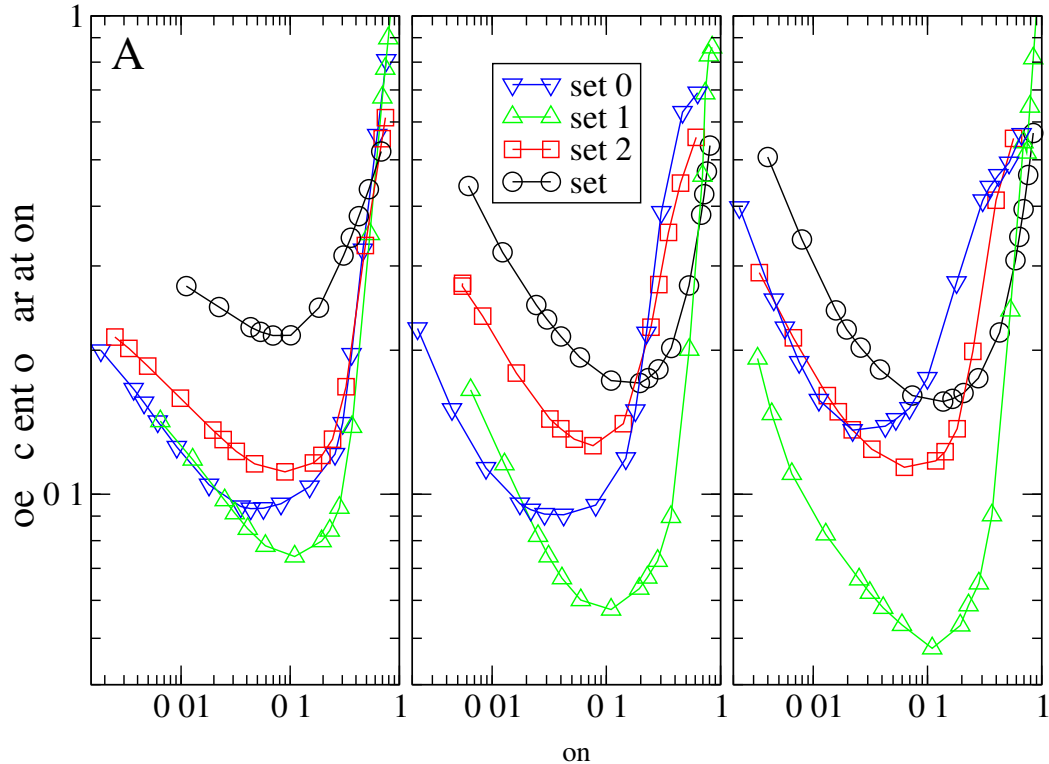}}
\caption{Coefficient of variation of the oscillations of the protein $P_2$, as function of the fraction of time that the gene spends in the active state, for 4 different sets of parameters (see the Supplementary Material for details) in systems with different degrees of cooperativity. A: $n=1$, B: $n=2$, C: $n=3$. $k_1$ satisfies $k_1=\frac{k_2 k_6 E_0}{k_3 p_{on}}$, and $k_{off}=f(k_1)$ (Eq. (\ref{eq.koff})). The rest of the parameters used correspond to the sets indicated in the legend. 
} \label{figure4}
\end{figure}

It is clear from Fig. \ref{figure2} that the differences in the mRNA temporal profile are closely related to the regularity of the oscillations of the protein. There are many ways to quantify this regularity. We have chosen to study the distribution of the time intervals between the minima of the protein abundance, since a sharply peaked distribution should signal the presence of very regular oscillations (the computation of the minima is explained in the Supplementary Material). To quantify this, we have calculated the Coefficient of Variation (CV) of the distributions, which is defined as the quotient between the square root of the variance and the mean of the distribution. The results are shown in Fig. \ref{figure4} for genetic oscillators with three different degrees of cooperativity. As a function of $p_{on}$, the curve for CV has the same general form (a curve with one minimum), in all the cases we have analyzed. To show that this feature is not an artifact of the method we have chosen to characterize the oscillations, we have tried two other different methods. In the first method we have calculated the CV of the period given by the Fourier spectrum of the time series of the protein abundance, and in the second we have calculated the relative difference between the periods obtained using the protein zeros and the Fourier spectrum. For both methods the results, given in the Supplementary Material, were qualitatively the same: oscillations are most regular for values of $p_{on}$ very similar to those in Fig. \ref{figure4}.

\begin{figure} 
\centerline{\includegraphics[width=\columnwidth,clip=true]{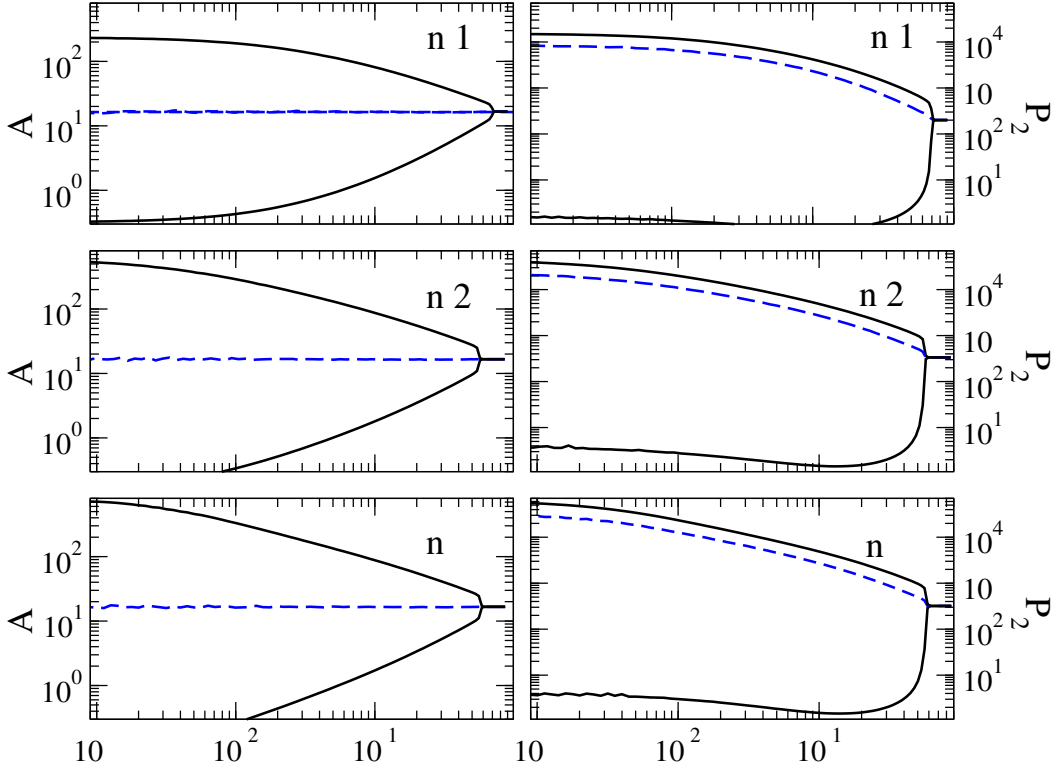}}
\caption{Bifurcation diagrams for $M(t)$ (left column) and $P_2(t)$ (right column) of the deterministic system, for three different cooperativities, as a function of $p_{on}$. $k_1$ satisfies $k_1=\frac{k_2 k_6 E_0}{k_3 p_{on}}$, and $k_{off}=f(k_1)$ (Eq. (\ref{eq.koff})). The rest of the parameters used correspond to set 2. The upper and lower black lines represent the values of the maximum and minimum, respectively, of the oscillation of the corresponding quantity. The dotted line represents the time average.} \label{figure5}
\end{figure}

Fig. \ref{figure5} shows that the values of $p_{on}$ for which the stochastic oscillations are most regular correspond to the parameter region where the deterministic system oscillates, when all the other parameters are given by set 2. We have checked that the same happens for all other parameter sets. Interestingly, we find that for arbitrarily small values of $p_{on}$ the system always displays deterministic oscillations with an amplitude inversely proportional to $k_1$, but the corresponding stochastic oscillations become less regular (see also the lower panels of Fig. \ref{figure2}). Thus, this behavior cannot be associated to the excitability of the deterministic system, as happens in other cases where the stochastic oscillations are irregular~\cite{Gonze2002,Gonze2004}, because it happens in a parameter region where the deterministic system not only oscillates but it is also far from the Hopf bifurcation.

It is not difficult to understand why oscillations are less robust when $p_{on}$ is relatively large. In this case the gene is active most of the time and, as the rate at which mRNA is transcribed ($k_1$) is small, the fluctuations in the number of molecules of mRNA tend to be large. The reason why small values of $p_{on}$ make oscillations less robust is perhaps less clear. Fig.~\ref{figure2} shows that the problem is that even though decreasing $p_{on}$ gives shorter and more intense bursts, the fluctuations tend to be rather large. In order to assess the generality of this behavior, in the next section we use some approximations to calculate explicitly the fluctuations of some variables for short bursts.

\section{Case of short bursts}
\label{shorts}

As was shown above, when $p_{on}$ is small, the system is most of the time in the regime where the degradation path of $P_2$ is heavily saturated. As a consequence, degradation is relatively slow and the time between successive activations of the gene becomes proportional to the height of the maximum of $P_2$ during that interval. This quantity, in turn, is proportional to the amount of mRNA generated during the preceding burst. When the gene is active mRNA is generated at a rate $k_1$. Thus, if a burst has a duration $T$, the probability $p(M|T)$ that $M$ copies of mRNA are produced during the burst is a Poisson distribution with event rate $k_1 T$. The coefficient of variation of the number of copies of mRNA generated during one burst can then be approximated by
\begin{eqnarray}
(CV_{mRNA})^2 &=& \frac{\int_0^{\infty} \sum_{M=0}^{\infty} M^2 p(M|T) p(T) dT}{(\int_0^{\infty} \sum_{M=0}^{\infty}  M p(M|T) p(T) dT)^2}-1 \nonumber \\
&=& \frac{ \int_0^{\infty} [k_1 T + (k_1 T)^2] p(T) dT}{(\int_0^{\infty} k_1 T  p(T) dT)^2}-1 \nonumber \\
&=& (CV_T)^2+\langle k_1 T \rangle^{-1}
\label{eq.CV}   
\end{eqnarray}

\noindent where $p(T)$ is the distribution function of the duration of the bursts. As was to be expected, the coefficient of variation of mRNA copies is closely related to the CV of the duration of the bursts.

To obtain an analytical expression for $p(T)$ it is necessary to make some approximations. We will only consider those bursts that are produced when there are no $P_2$ molecules present, because this produces the longest bursts, which in turn generate most of the mRNA present in the cell, when $k_1$ is large. If the unbinding of $P_2$ from the activator is relatively fast, we can assume that the burst begins with the gene in state $D_0$. Thus, the gene only becomes repressed again after the binding of $n$ molecules to the activator. We assume that the binding of $P_2$ molecules happens at a rate $k_{off} P_2$, and that the rate of unbinding is much smaller, so that unbinding can be neglected. The length of the burst is thus a random variable that can be associated to the arrival time of the $n$-th $P_2$ molecule. If the rate $k_{off} P_2$, which is a random variable, is approximated by $k_{off} P_2(t)$, we obtain a non-homogeneous Poisson process~\cite{Gallager}, and $p(T)$ is the corresponding probability density function of the $n$-th arrival time:
\begin{equation}
p(T)=\dfrac{k_{off}^n}{(n-1)!} \left( \int_0^{T} P_2(t) dt \right)^{n-1} P_2(T) \exp \left(-k_{off} \int_0^{T} P_2(t) dt \right)
\label{eq.pT}   
\end{equation}

\begin{figure} 
\centerline{\includegraphics[width=0.7 \columnwidth,clip=true]{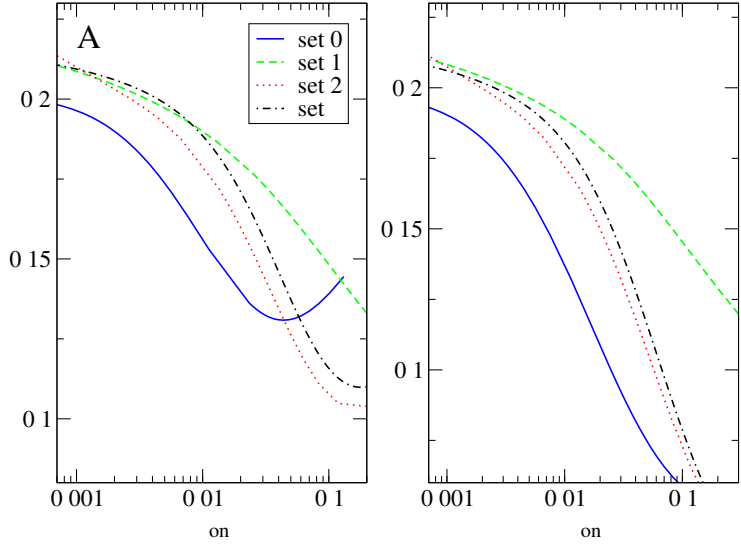}}
\caption{Deterministic approximation for the coefficient of variation of the number of mRNA copies (panel A) and the length of the bursts (panel B) for the system with $n=3$. The sets of parameters used are the same as in the previous figures.
} \label{figure6}
\end{figure}

To estimate the average abundance of $P_2$ during a burst, we assume that its average $P_2(t)$ follows the evolution given by Eqs. (\ref{eq.detailed}), but with the restriction $R \equiv 0$. Solving these equations for the same sets analyzed in the previous sections, and using Eqs. (\ref{eq.CV}) and (\ref{eq.pT}) we obtain an approximation for the coefficient of variation for the length of a burst and for the number of mRNA copies produced. Fig. \ref{figure6} shows that the behavior of these quantities is qualitatively similar to that observed for the coefficient of variation of the  period in the fully stochastic system (Fig. \ref{figure4}). Notice that we are dealing here with only one of the sources of randomness. For example, we have only used one initial condition for the equations, i.e. only one initial value for the abundances of proteins and mRNA at the beginning of the burst, even though they are stochastic variables. Furthermore, we are not considering the fluctuations of events happening in the interval between bursts.

\begin{figure} 
\centerline{\includegraphics[width=\columnwidth,clip=true]{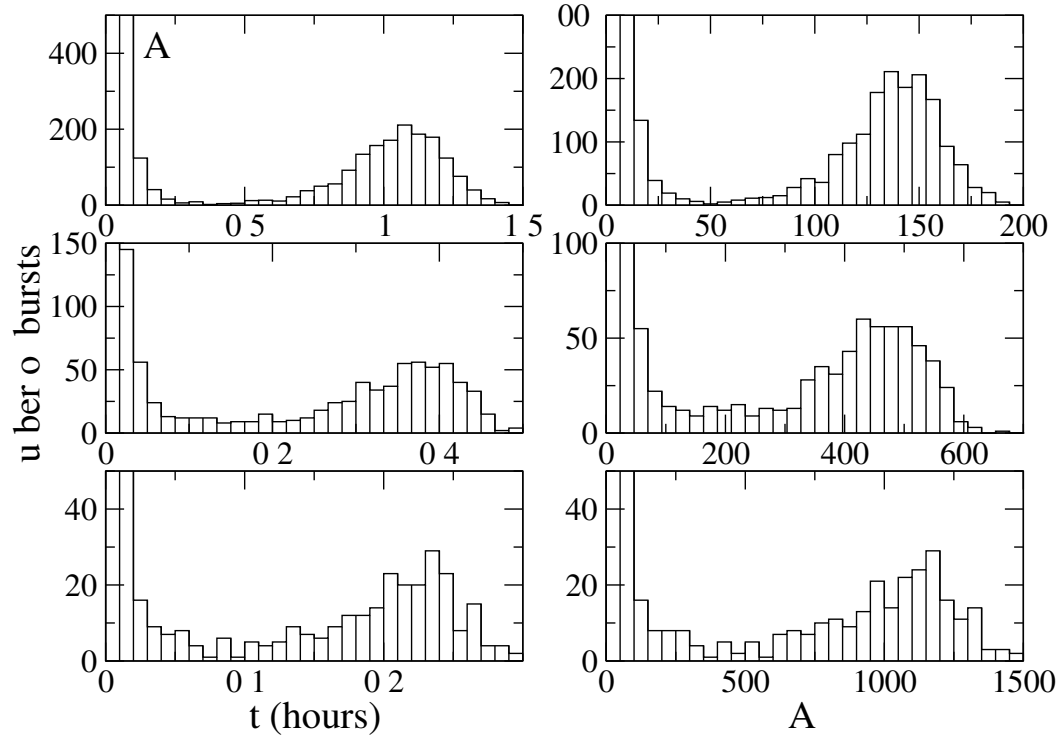}}
\caption{Histograms of the bursts generated in three different simulations of 35000 system hours of a system with $n=3$, as a function of the duration of the bursts (left column) or of the number of mRNA molecules produced in the burst, for three values of $p_{on}$. A,B: $p_{on}=0.063$; C,D: $p_{on}=0.0067$; E,F: $p_{on}=0.0017$. The parameters used correspond to set 2 (see Supplementary Material). The length of all simulations was 35000 hours.
} \label{figure7}
\end{figure}

Even though the approximations we use are rather crude, they seem to capture what happens in the full system. Fig. \ref{figure7} shows that both the distribution of duration of the bursts and the distribution of mRNA molecules produced tend to get wider as $p_{on}$ increases (in the sense that the average decreases faster than the variance).

\section{Long bursts and stability of the deterministic system}
\label{longs}

In the deterministic analogues of the stochastic systems analyzed above, there is evidently no counterpart to the quality of oscillations because, by definition, deterministic oscillations are always regular. But we can study what happens to the {\it stability} of the equations when $k_{off}$ and $k_1$ are modified following Eq. (\ref{eq.koff}). For this we first turn to a quasi-steady-state approximation~\cite{Segel}, where we assume that the interactions with the enzyme are much faster than the other interactions involved. Within this approximation, the equations for $M(t)$, $P_1(t)$ and $P_2(t)$ are
\begin{eqnarray}
\dot{M} &=& k_1 D \frac{1+\alpha_7 P_2+\alpha_7 \alpha_8 P_2^2}{1+\alpha_7 P_2+\alpha_7 \alpha_8 P_2^2+\alpha_7 \alpha_8 (k_{on}/k_{off}) P_2^3}  - k_2 M,\nonumber \\
\dot{P_1} &=& k_3 M - k_4 P_1, \nonumber \\ 
\dot{P_2} &=& k_4 P_1 -k_6 E_0\frac{P_2}{x+P_2} 
\label{eq.QSSA2}
\end{eqnarray}

\begin{figure} 
\centerline{\includegraphics[width=\columnwidth,clip=true]{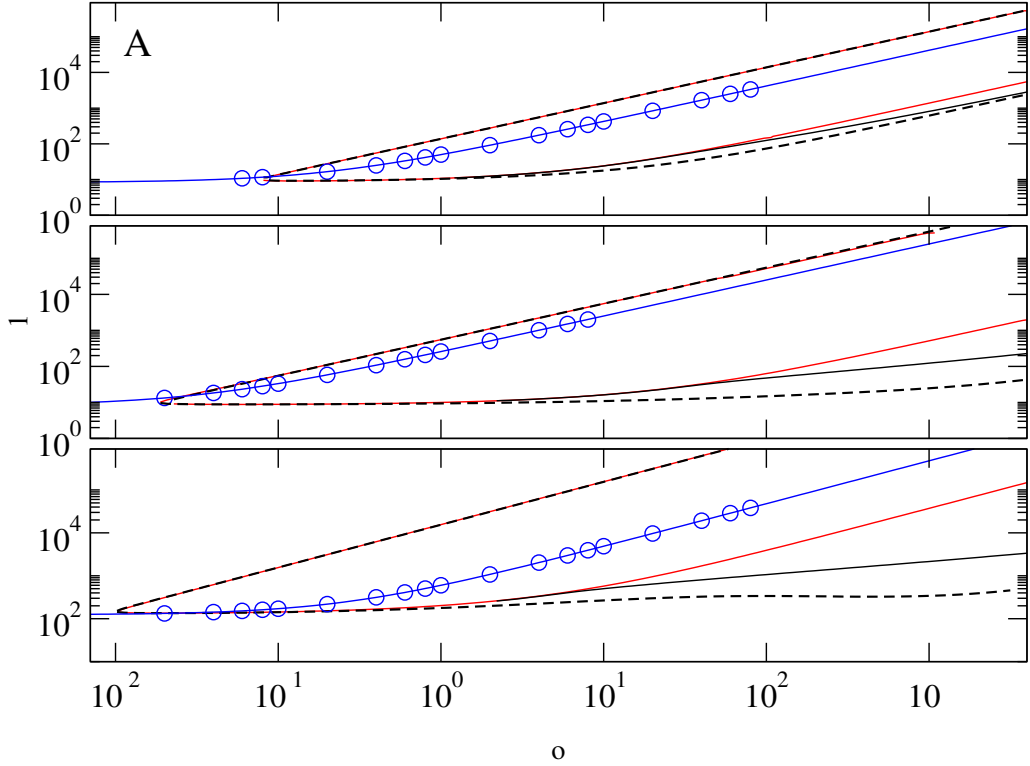}}
\caption{Asymptotic stability for deterministic genetic oscillators with three different cooperativities. In the region between the dashed curves the equilibrium is unstable and there is a limit cycle. Outside this region the equilibrium is asymptotically stable. In the region between the full red curves the reduced version of the system is unstable and has a limit cycle. In the region between the lower red line and the black line the system is bistable. The symbols represent the parameters values used in Fig.~\ref{figure4}. A: $n=1$, parameters corresponding to set 2. B: $n=2$, parameters corresponding to set 2. C: $n=3$, parameters corresponding to set 1.
} \label{figure8}
\end{figure}

Fig. \ref{figure8} shows the phase diagram, in the plane of the parameters $k_1$ and $k_{off}$, for both the full deterministic system and the reduced system, for $n=1$, $n=2$ and $n=3$. The common feature in all cases is that, for fixed values of $k_{off}$, there is an interval of values of $k_1$ where the equilibrium is unstable and the variables display oscillations, provided that $k_{off}$ is larger than a threshold value. The curve that gives the upper limit of this interval is the same for the full and the reduced system. In both cases the curve represents the appearance of a Hopf bifurcation. On the other hand, the curves that give the lower limit of the interval represent different kind of transitions: the detailed system has a second Hopf bifurcation, whereas for the reduced system the limit cycle appears for values of $k_1$ that are smaller than the value that destabilizes the equilibrium. This lower curve (a full black line in Fig. \ref{figure7}), obtained by numerical simulations, is likely to represent a fold limit cycle bifurcation~\cite{Izhikevich}.

We can calculate analytically the stability of the reduced system using the Routh-Hurwitz criterion~\cite{Murray}. For the case of $n=3$, we obtain
\begin{equation}
(k_2+k_4)(\gamma_2(k_2+k_4) + k_2 k_4 + \gamma_2^2)>k_1 k_3 k_4 \alpha_7 \alpha_8 \alpha_o (P_2^*)^2 \beta f^2(k_{off})
\label{eq.RW}
\end{equation}

\noindent where  
\begin{eqnarray}
\gamma_2 &=& \frac{k_6 E_0 x}{(P_2^*+x)^2} \nonumber \\
\beta &=& \frac{3+2 \alpha_7 P_2^* + \alpha_7 \alpha_8 (P_2^*)^2}{(1+ \alpha_7 P_2^* + \alpha_7 \alpha_8 (P_2^*)^2)^2}
\label{eq.g2D}
\end{eqnarray}
\noindent and $P_2^*$ is the value of $P_2$ at the fixed point of Eqs. \ref{eq.QSSA2}.

In the regime we consider in this paper, where $k_{off}$ and $k_1$ are related by Eq. (\ref{eq.koff}), in order to modify the temporal profile of mRNA expression but keeping constant the values of $M$, $P_1$ and $P_2$ at equilibrium, the right term of Eq. (\ref{eq.RW}) becomes an increasing function of $k_{off}$, whereas the left term is a positive constant. This implies that if, when all other parameters are kept fixed, the system is in the oscillation region for some value of $k_{off}$ and $k_1$, then it will also oscillate for all larger values of $k_{off}$. Using the same reasoning, it is evident that the system should leave the oscillation region for small enough values of $k_{off}$. Thus, within this approximation, there is always a threshold value of $k_{off}$ below which the deterministic system is asymptotically stable. This can be easily generalized to all values of $n$. For the complete deterministic system, given by Eq. (\ref{eq.detailed}), the calculations are much less straightforward, but using symbolic mathematics software we have been able to confirm that the same behaviour is present in the detailed system for $n=1$, $n=2$ and $n=3$. Fig. \ref{figure8} shows an example of three of the regimes analyzed in the previous sections. It can be observed that when $k_{off}$ is small enough the curves leave the instability region.

Turning now to the stochastic system, the reasoning above leads us to expect that, when the bursts are long (i.e. when $k_{off}$ is small) the quality of the oscillations decreases, as the deterministic system is close to the Hopf bifurcation. This has been verified in many systems (see e.g. ~\cite{Gonze}) and it happens because the amplitude of the oscillation becomes smaller as the system gets closer to the Hopf bifurcation and the effects of stochastic noise become more important. When the system is below the Hopf bifurcation there can be stochastic oscillations without deterministic oscillations (i.e. the oscillations are induced by the noise) but it is likely that the quality of the oscillations becomes small for systems that are far from the bifurcation (in analogy with the case of systems with stable foci where the quality is smaller when the foci are more stable~\cite{McKane}).

Interestingly, it is possible to find regimes for which the threshold value of $k_{off}$, and therefore of $k_1$, is relatively large, which implies a regime where deterministic oscillations can only occur if bursts are both very short and intense. But in general the curves that correspond to this regime are very close to the bifurcation curves, and thus the corresponding oscillations are very irregular. Therefore, the regimes where intense bursting is necessary for oscillations do not seem to be biologically relevant.

\section{Discussion}

Transcriptional bursting has been shown to be a widespread phenomenon in gene expression, but its causes and function are still a hot topic of research. The case of genes expressing proteins that act as genetic oscillators is qualitatively different from those associated to proteins that are constitutively expressed. The fundamental difference is that in genetic oscillators bursts can be due to the interaction between the activator of the gene and the very protein associated to that gene. In this paper we have studied the relationship between such transcriptional bursts and the stochastic oscillations of the protein.

Given that genetic oscillators are the building blocks of various behavioural rhythms, we have concentrated on determining which are the transcriptional regimes that lead to very regular protein oscillations. For this we have quantified the bursting regime using $p_{on}$, which represents the fraction of time that the gen is active. To focus only on the temporal aspect of bursting we have studied the regime where some parameters of the system are varied, but the average amount of mRNA produced is kept fixed.

Using numerical simulations for several different sets of parameters we have shown, for genetic oscillators with three different levels of cooperativity, that there is an amount (and an associated intensity) of bursting which is the best in terms of the quality of the stochastic oscillations produced. The best value for $p_{on}$ is in most cases $p_{on} \approx 0.1$, which means that the gene is in the active state approximately $10 \%$ of the time. Interestingly, this value is similar to what has been found for many genes, some of them even encoding clock proteins~\cite{Suter}. Our result implies that long, low-intensity, bursts tend to give very irregular fluctuations, but also that the same happens if the bursts are too short and intense, which is much less intuitively obvious. Using some approximations, we show that this can be at least partially explained by the fluctuations in mRNA copies and burst duration. More specifically, we have shown that what happens is that when $p_{on}$ decreases, both the variance and the average of the distribution of burst length, and mRNA produced, decrease, but the variance decreases more slowly, thus producing increasingly large fluctuations. Both the simulations and the approximation show that this effect becomes more significant when more steps of cooperativity are added.

The fact that the best quality of the oscillations is reached at a finite value $p_{on}$ is interesting because this is not what happens when other possibilities for transcription control are considered. For example, it has been shown~\cite{Forger,Gonze,Nishino} that if the rate of binding and unbinding are multiplied by the same constant, the quality of the stochastic oscillations is a monotone increasing function of this constant. 

The relationship between burstiness and oscillations has also been studied in detail for systems where the oscillation only appear because of the stochastic noise (i.e. noise-induced oscillations)~\cite{Toner}. There it was found that the influence of burstiness in the quality of oscillations is monotonous, but depending on the system, the quality may be an increasing or a decreasing function of burstiness. Because they provide a fully analytical treatment of many models, the authors consider only models with two species. It would be interesting to see whether for more complex systems the quality can be a non-monotonous function of the burstiness, as in the model analyzed here.

The analysis we provide in this paper is mainly numerical, supplemented with two theoretical approximations for very short and very long bursts, which allows us to suggest that the nonmonotonicity of the quality factor may be the general case for this system. To go beyond this, one should probably perform a theoretical analysis of the effects of perturbations of the limit cycle, along the lines Ref.~\cite{Gaspard}. The problem is that the system is much more complicated and has some variables (those related with the gene) that can only take values 0 and 1, which should therefore be given a different treatment in a system size expansion.

We have also shown here that in the case of genes encoding clock (i.e. oscillating proteins), the distribution of burst duration is not necessarily different to the ones observed for genes associated to constitutively expressed proteins. Differences only appear if repression of the gene activators is caused by many copies of the protein. 

\section*{Acknowledgements}

SR-G acknowledges financial support from ConICET through project PIP 11220120100495.

\end{document}


\begin{frontmatter}



\title{Supplementary Material for \\Role of transcriptional bursts in cellular oscillations}


\author{N. Almeira}
\address{Facultad de Matem{\'a}tica, Astronom{\'\i}a, F{\'\i}sica y Computaci{\'o}n, Universidad Nacional de
C{\'o}rdoba, Ciudad Universitaria, C{\'o}rdoba, Argentina.\\
Instituto de F{\'\i}sica Enrique Gaviola (IFEG-CONICET), Ciudad Universitaria, 5000 C{\'o}rdoba,
Argentina.}

\author{S. Risau-Gusman}
\address{Centro At{\'o}mico Bariloche,
8400 S. C. de Bariloche, Argentina \\
Consejo Nacional de Investigaciones Cient{\'\i}ficas y
T{\'e}cnicas, Argentina}

\begin{keyword}

\PACS
\end{keyword}
\end{frontmatter}

\section{Parameters used in the simulations}
\label{Intro}

For the paper we have performed simulations for three systems of genetic oscillators where it takes $n$ molecules of the protein to repress the activator of the gene. We have simulated systems with $n=1$, $n=2$, and $n=3$, and in each case we have tested four different sets of parameters. The values of these sets are given in Table 1.

\begin{table}[!htbp]
\begin{tabular}{|c|c|c|c|c|c|c|c|c|c|c|c|c|}
\cline{3-13}
 \multicolumn{2}{c|}{} & $k_2$ & $k_3$ & $k_4$ & $k_5$ & $k_{-5}$ & $k_6$ &  $k_{on}$ & $k_7$ & $k_{-7}$ & $k_8$ & $k_{-8}$ \\ \hline
\multirow{4}{*}{$n=1$} 
 & set 0 &  0.5 & 200. & 0.5 & 2. & 0.01 & 5. & 40. & - & - & - & - \\ \cline{2-13}
 & set 1 &  0.5 & 20. & 0.5 & 2. & 0.01 & 5. &  40. & - & - & - & - \\ \cline{2-13}
 & set 2 &  0.5 & 40. & 0.5 & 2. & 0.01 & 0.666 &  40. & - & - & - & - \\ \cline{2-13}
 & set 3 &  1 & 20. & 2 & 2. & 0.01 & 0.5 &  40. & - & - &- & - \\ \hline
\multirow{4}{*}{$n=2$} 
& set 0 &  0.5 & 200. & 0.5 & 2. & 0.01 & 5. & 40. & 1. & 40. & - & - \\ \cline{2-13}
 & set 1 &   0.5 & 20. & 0.5 & 2. & 0.01 & 5. & 40. & 1. & 40. & - & - \\ \cline{2-13}
 & set 2 &  0.5 & 40. & 0.5 & 2. & 0.01 & 0.666 & 10. & 1. & 40. & - & - \\ \cline{2-13}
 & set 3 &   1 & 40. & 2 & 2. & 0.01 & 0.5 & 5 & 1. & 40. & - & - \\ \hline
\multirow{4}{*}{$n=3$} 
& set 0 &  0.5 & 300. & 2 & 2. & 0.1 & 5. & 40. & 1. & 20. & 1. & 40 \\ \cline{2-13}
& set 1 &  0.5 & 20. & 0.5 & 2. & 0.01 & 5. & 40. & 1. & 40. & 1. & 40 \\ \cline{2-13}
& set 2 &  0.5 & 40. & 0.5 & 2. & 0.01 & 0.666 &  40. & 1. & 40. & 1. & 40 \\ \cline{2-13}
& set 3 &  1   & 40. & 2 & 2. & 0.01 & 0.5 & 5 & 1. & 40. & 1. & 40 \\
\hline
\end{tabular}
\caption{Values of the parameters used in the simulations of the three genetic oscillators, with cooperativities $n=1$, $n=2$ and $n=3$, mentioned in the main text.}
    \label{tab:tab1}   
\end{table}

For each set of parameters we have also determined the equilibrium values of mRNA ($M^*$), the two states of the protein ($P_1^*$ and $P_2^*$), and the free enzyme ($E^*$). Using these values, given in Table 2, and Eq. (6) of the main text, for every value of $k_{off}$ one can obtain the corresponding value of $k_1$

\begin{table}[!htbp]
\begin{tabular}{|c|c|c|c|c|c|}
\cline{3-6}
 \multicolumn{2}{c|}{} &  $M^*$ & $P_1^*$ & $P_2^*$ & $E^*$\\ \hline
\multirow{4}{*}{$n=1$} 
 & set 0 &   24.94 & 9976.94 & 1084.06 & 2.30  \\ \cline{2-6}
 & set 1 &   246.03. & 9841.06 & 155.10 & 15.89  \\ \cline{2-6}
 & set 2 &   16.63 & 1331.09 & 200.41 & 1.69 \\ \cline{2-6}
 & set 3 &   12.46 & 249.28 & 88.37 &  2.88  \\ \hline
\multirow{4}{*}{$n=2$} 
& set 0  & 24.97 & 9989.01 & 2276.58 & 1.1 \\ \cline{2-6}
 & set 1 & 246.70 & 9868.15 & 187.48 & 13.18 \\ \cline{2-6}
 & set 2 & 16.65 & 1331.99 & 336.15 & 1.00 \\ \cline{2-6}
 & set 3 & 12.42 & 248.42 & 40.04 & 6.33 \\ \hline
\multirow{4}{*}{$n=3$} 
& set 0 & 16.62 & 2493.11 & 921.78 & 2.76 \\ \cline{2-6}
& set 1 & 246.8 & 9871.95 & 193.12 & 12.80 \\ \cline{2-6}
& set 2 & 16.63 & 1330.66 & 336.17 & 1.00  \\ \cline{2-6}
& set 3 & 12.45 & 248.99 & 63.28 & 4.01  \\
\hline
\end{tabular}
\caption{Abundances of mRNA, the two states of protein, and the free enzyme, at the equilibrium given by the parameter sets given in Table 1.}
    \label{tab:tab2}   
\end{table}

\newpage

\section{Characterization of the oscillations}
\label{characterization}

As explained in the main text, to characterize the stochastic oscillations of the abundance of the protein $P_2$ we have chosen to analyze the distribution of time intervals between "minima", loosely defined as the times when the protein copy number reaches its smallest values. The problem with this definition is that when the number of copies of the protein is very low, very rapid fluctuations dominate the dynamics until the number of copies begins to raise again. Thus, we have resorted to the following simple approximation to estimate the position of the minima we are really interested in. First we have recorded the times $t_i^+$ and $t_i^-$ at which $P_2$ crosses the line $P_2=N_{P2}$, coming from above or below that value, respectively. Then we have eliminated very rapid fluctuations by discarding consecutive times with a difference of less than 1 minute. The position of the minima were then approximated as $(t_i^{+}+t_i^-)/2$. We have typically used $N_{P2}=100$ as a threshold, but we have checked that the distributions are practically the same if the threshold is between $50$ and $300$. In order to see whether the distributions of time intervals between minima obtained are reasonable, we have compared them with the Fourier transform of $P_2$. We have checked that in general the peaks of both functions are very close, as Fig. 1 shows.
 
\begin{figure} 
\centerline{\includegraphics[width=\columnwidth,clip=true]{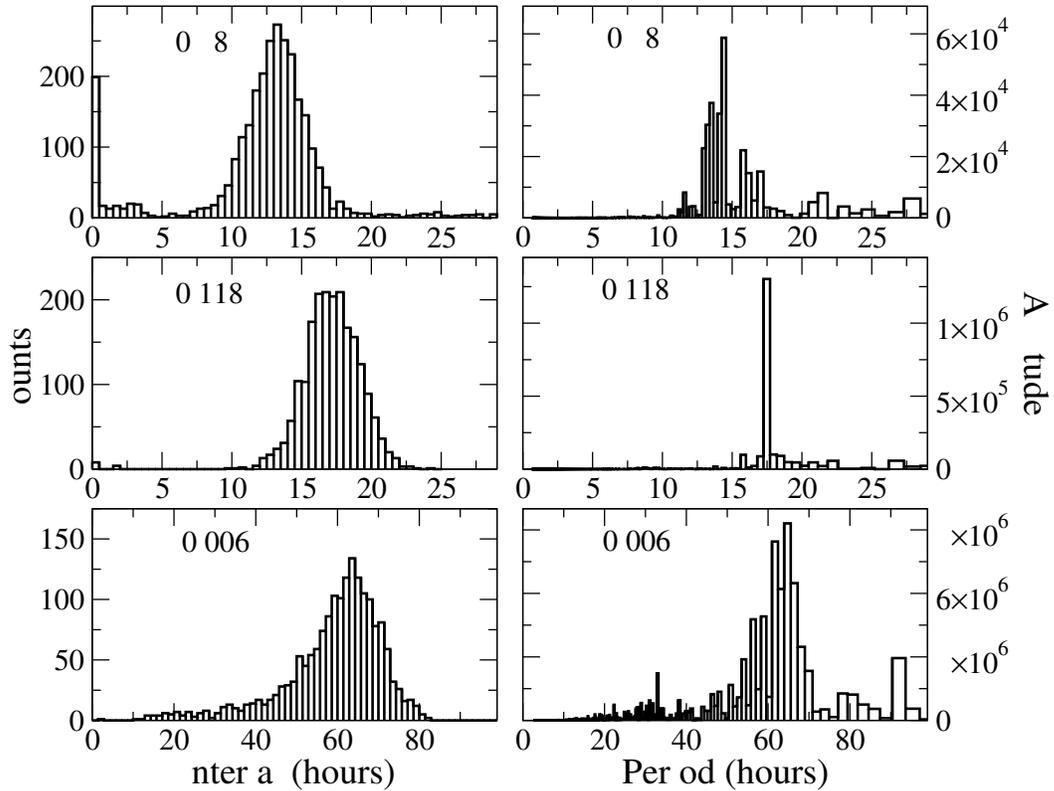}}
\caption{Distributions of time intervals between minima of $P_2(t)$ (panels on the left) and discrete Fourier transforms of $P_2(t)$ (panels on the right) for three different values of $p_{ON}$, for a system with $n=2$. The other parameters are as in Set 2 (see Table 1). For the two upper rows 
we simulated 35000 hours of the system, and for the lowest row of panels we simulated 105000 hours of the system.
} \label{figure1}
\end{figure}

In order to check that the results of the main text are not an artifact of the method used to characterize the oscillations, we have used two additional methods. In the first one we have performed simulations for each of the tested systems using at least 200 different initial conditions and we have calculated the period of the oscillation in each case, using the discrete Fourier transform. Then we have calculated the CV of the distribution of the periods obtained. The results are shown in Fig. 2.

\begin{figure} 
\centerline{\includegraphics[width=\columnwidth,clip=true]{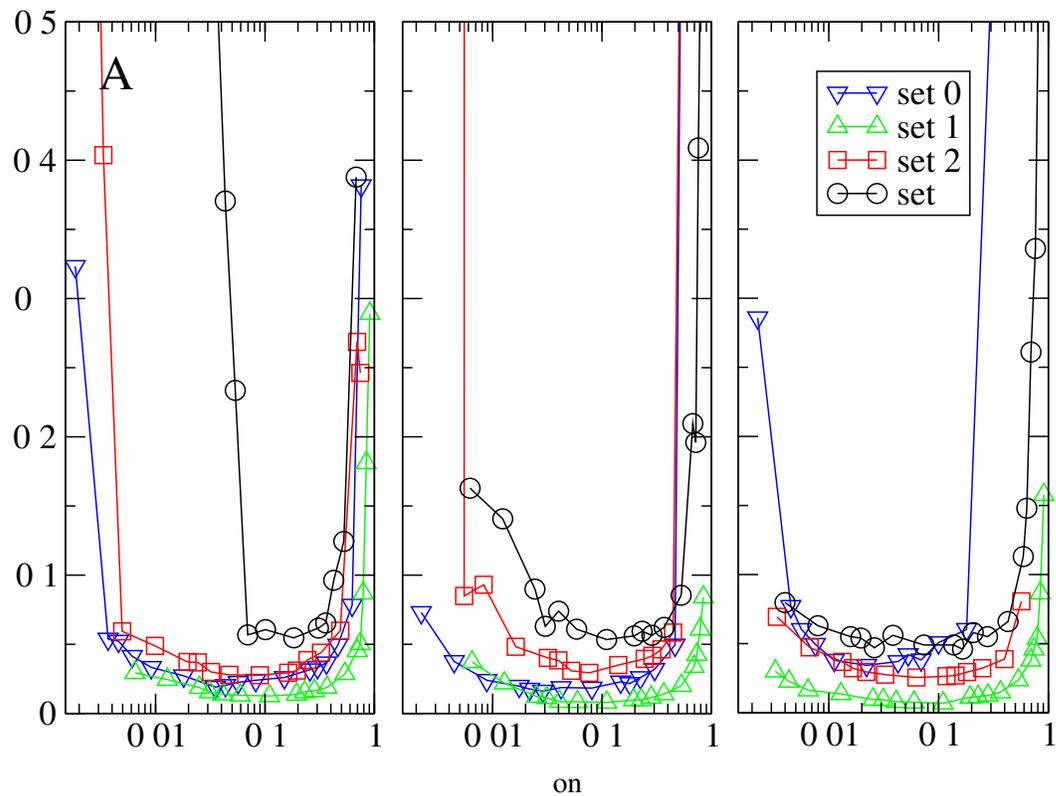}}
\caption{Coefficient of variation of the discrete Fourier transform of the abundance of the protein $P_2$, as function of the fraction of time that the gene spends in the active state, for 4 different sets of parameters (see Table 1) in systems with different degrees of cooperativity. A: $n=1$, B: $n=2$, C: $n=3$.
} \label{figure2}
\end{figure}

In the second method we have characterized the oscillations by the relative difference of the period $P_F$ obtained by taking the discrete Fourier transform of the protein time series, and the period $P_z$ given by the average of the distribution of distance between consecutive minima of the protein. Because of finite size effects the abundance does not vanish at all these minima. Fig. 3 shows the values obtained for the quantity $\Delta=2|P_F-P_z|/(P_F+P_z)$, for each set.

\begin{figure} 
\centerline{\includegraphics[width=\columnwidth,clip=true]{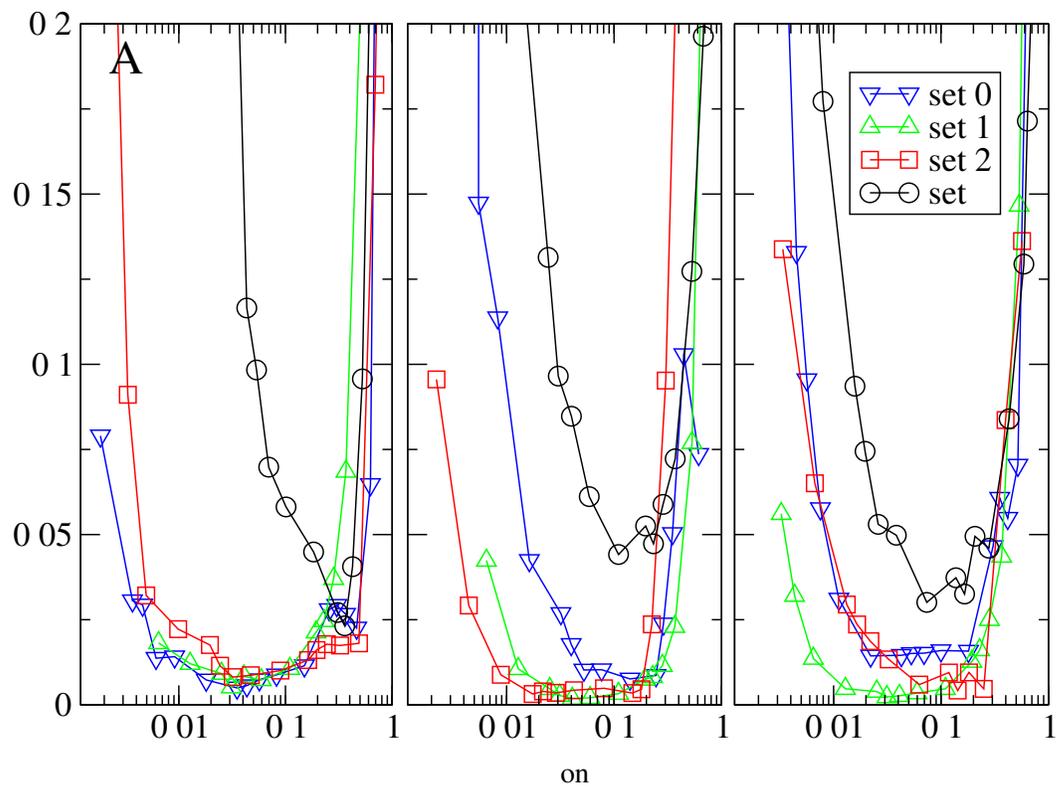}}
\caption{Relative difference between the periods obtained using a Fourier transform and the distribution of the distance of minima of the abundances of the protein $P_2$, as function of the fraction of time that the gene spends in the active state, for 4 different sets of parameters (see Table 1) in systems with different degrees of cooperativity. A: $n=1$, B: $n=2$, C: $n=3$.} \label{figure3}
\end{figure}